\documentclass[12pt,a4paper,final]{iopart}

\usepackage{iopams}  
\usepackage{graphicx}

\usepackage[breaklinks=true,colorlinks=true,linkcolor=blue,urlcolor=blue,citecolor=blue]{hyperref}

\begin{document}

\title[Non-adiabatic charge state transitions in singlet-triplet qubits]{Non-adiabatic charge state transitions in singlet-triplet qubits}

\author{Tuukka Hiltunen, Juha Ritala, Topi Siro, and Ari harju}
\address{ Department of Applied Physics and Helsinki Institute of Physics,
  Aalto University School of Science, P.O. Box 11100, 00076 Aalto, Finland}

\begin{abstract}
In double quantum dot singlet-triplet qubits, the exchange interaction is used in both quantum gate operation and
the measurement of the state of the qubit. The exchange can be controlled electronically by applying gate voltage pulses.
We simulate  the exchange induced charge state transitions in one and two singlet-triplet qubit systems using the exact diagonalization method.
We find that fast detuning pulses may result in leakage between different singlet charge states. The leakage
could cause measurement errors and hinder quantum gate operation for example in the case of the two-qubit Coulomb gate.
\end{abstract}

\vspace{2pc}
\submitto{\NJP}

\section{Introduction}

The development of experimental methods
has enabled the fabrication of ``artificial atoms" 
with a controlled
number of electrons, ranging from a few to a few hundred, confined in
a tunable external potential inside a semiconductor ~\cite{Ashoori96,Reimann02,Saarikoski_RMP}. 
These quantum dots (QD's) have been
proposed as a possible realization for the qubit of a quantum
computer \cite{Loss98,Hanson07}.

A framework for using two-electron spin eigenstates as qubits was proposed by Levy in 2002 \cite{levy}.
The two-electron double quantum dot (DQD) spin states have natural protection against the decoherence by the hyperfine
interaction and allow for a scalable architecture for quantum computation \cite{taylor}.
The universal set of quantum gates for two spin singlet-triplet DQD qubits has been demonstrated
experimentally. These gates include one qubit rotations
generated by the exchange interaction \cite{petta2}, stabilized hyperfine magnetic field
gradients \cite{foletti}, and two qubit operations using long distance capacitative coupling by the Coulomb interaction \cite{weperen,shulman}.

The exchange interaction results from the symmetry properties of the spatial many-body wave function. In the singlet state,
the electrons behave effectively like bosons, an overlap of the wave functions of the electrons lowers the energy of the state. In the triplet state, the effect is opposite.
In singlet-triplet DQD qubits, the exchange interaction can be turned on by electrically detuning the two dots of the qubit by applying gate voltages \cite{petta2}.

In $S-T_0$ qubits, the exchange interaction is used to drive both one qubit rotations \cite{Loss98,Pettaa,petta2,dial,taylor2,foletti} and two-qubit 
interactions \cite{levy,li,weperen,shulman,hanson,stepa}. In addition to quantum gate operation, the exchange interaction is also exploited in measuring the state of the $S-T_0$ qubit \cite{petta2,barthel1}.
As the detuning of the dots is increased, the singlet state localizes into the dot with lower potential, undergoing a transition from $(1,1)$ charge state
(one electron in each dot) to $(0,2)$. Due to the repulsive exchange force, the triplet stays in $(1,1)$ \cite{petta2,taylor2,yjunc}. The state of the qubit can then
be measured using a charge sensor.

In this paper, we use the exact diagonalization (ED) method to simulate the transition between singly and doubly occupied singlet states. The transition is induced by increasing the detuning between the dots of the qubit.
We study the effect of the speed of the detuning sweep and find that a fast increase can lead to Landau-Zener type leakage between the charge states.
We propose that this kind of leakage could cause errors in measuring the singlet probability. We also study the operation of the capacitatively coupling
Coulomb gate and discover that the leakage may result in the gate not achieving maximal Bell-state entanglement.

\section{Model and methods}

We model lateral GaAs quantum dot systems with the two-dimensional Hamiltonian
\begin{equation}
\label{eq:Hamiltonian}
H(t)=\sum_{j=1}^N\left[-\frac{\hbar^2}{2m^*}\nabla_j^2+V(\mathbf{r}_j,t)\right]+\sum_{j<k}\frac{e^2}{4\pi\epsilon r_{jk}},
\end{equation}
where $N$ is the number of electrons, $V$ the external potential, and $m^*\approx0.067\,m_e$ and
$\epsilon\approx12.7\,\epsilon_0$ are the effective electron mass and
permittivity in GaAs, respectively. In numerical work, it is
convenient to switch into effective atomic units by setting
$m^*=e=\hbar=1/4\pi \epsilon=1$. In these units, energy is given by
$\mathrm{Ha}^* \approx 11.30$~meV and length in $a_0^* \approx 10.03$~nm.

In our computations, a singlet-triplet qubit is modeled with a double quantum dot (DQD) potential. A system of two singlet-triplet qubits
is modeled as four quantum dots.
In the model, the external potential $V(\mathbf{r})$ for quantum dot systems
consists of several parabolic wells. A confinement potential of $M$ parabolic wells can be written as
\begin{equation}\label{eq:qdpot}
V(\mathbf{r},t)=\frac{1}{2}m^*\omega_0^2\min_{1\leq j \leq M}\{|\mathbf{r}-\mathbf{R}_j|^2\}+V_d(t,\mathbf{r}),
\end{equation}
where $\{\mathbf{R}_j\}_{1\leq j\leq M}$ are the locations of the minima of the parabolic wells and $\omega_0$ is the
confinement strength. A time dependent detuning potential $V_d(t,\mathbf{r})$ is included. 

\begin{figure}[!htb]
\vspace{0.3cm}
\includegraphics[width=0.5\columnwidth]{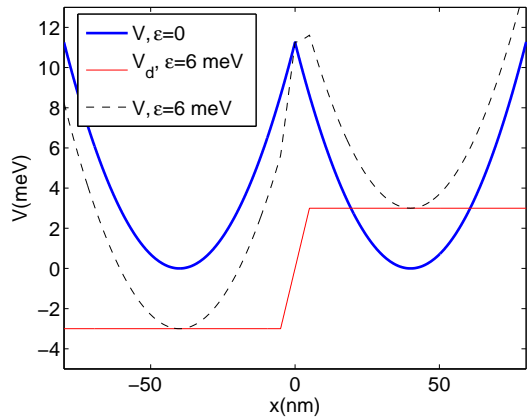}
\centering
\caption{A two dot potential and the detuning. The potential of Eq. (\ref{eq:qdpot}), with two minima at $\mathbf{R}_1=(-40$ nm$,0)$
and $\mathbf{R}_2=(40$ nm,$0)$ and with no detuning is shown in the $x$-axis (the thick blue line). The confinement strength is $\hbar\omega_0=4$ meV.
A detuning potential $V_d$ of $\epsilon=6$ meV is shown as the thin red line. The detuning is a step function that is made continuous
with a linear ramp. The detuned potential is shown with the dashed black line.
}
\label{fig:detuning}
\end{figure}

The detuning is modeled as
a step function that assumes constant values at each well. 
The detuning of a singlet-triplet qubit is defined as the potential energy difference between the two parabolic minima of the qubit,
i.e. if the qubit consists of the minima at $\mathbf{R}_1$ and $\mathbf{R}_2$, the detuning is $\epsilon(t)=V(\mathbf{R}_1,t)-V(\mathbf{R}_2,t)$.
A DQD potential and the detuning
are illustrated in Fig. \ref{fig:detuning}.

The Hamiltonian (\ref{eq:Hamiltonian}) is diagonalized using the ED method. In the ED calculations, the one-particle basis is the eigenstates corresponding
to the confinement potential (\ref{eq:qdpot}). The one-particle eigenstates are computed using the multi-center Gaussian basis (the method is described in detail by Nielsen et al. \cite{requ}).
The matrix elements $V_{i,j}=\langle\phi_i|V(\mathbf{r})|\phi_j\rangle$ and $V_{i,j,k,l}=\langle\phi_i|\langle\phi_j|V_{int}|\phi_l\rangle|\phi_k\rangle$
can be computed analytically in the Gaussian basis. The matrix elements corresponding to the one-particle eigenstates are then computed
from the Gaussian elements by a basis change.

In the computation of the one-particle eigenstates, an evenly spaced grid of about 200 Gaussian functions (209 in the two dot case and 189 in the four dot case) was used.
The grid dimensions and the Gaussian widths were optimized and the convergence of the states was verified by comparing the energies to ones obtained with a much larger grid of around 2000 Gaussians.
We performed the basis change of the interaction matrix elements $V_{i,j,k,l}$ with an Nvidia Tesla C2070 graphics processing unit, which was programmed with CUDA \cite{CUDA}, a parallel programming model for Nvidia GPUs.

The time evolution of the wave function is computed by propagating the initial ground state $\psi(0)$,
\begin{equation}\label{eq:exp}
\psi(t+\Delta t)=\exp\left(-\frac{i}{\hbar}H(t)\Delta t\right)\psi(t).
\end{equation}
The ED Hamiltonian is stored as a sparse matrix and its eigenvalues and eigenvectors are computed by the Lanczos iteration.  The ground state of the system can be obtained by directly applying the Lanczos method. The higher states can then be
computed by a 'ladder operation'. The $n$th state $|\psi_n\rangle$ is obtained as the ground state of the Hamiltonian
\begin{equation}
H_n=H+\delta\sum_{j=1}^{n-1}|\psi_j\rangle\langle\psi_j|,
\end{equation}
where $H$ is the original Hamiltonian of the system, $\delta>0$ is a penalizing term and $\{|\psi\rangle_j\}_{j=1}^{n-1}$ are the eigenstates below the $n$th. 
The matrix exponentiation in
(\ref{eq:exp}) is also done by Lanczos.

\section{One qubit}

The logical basis of a two-electron singlet-triplet qubit consists of the two lowest eigenstates,
the singlet state, $|S\rangle=\frac{1}{\sqrt{2}}(|\uparrow\downarrow\rangle-|\downarrow\uparrow\rangle)$, 
and the $S_z=0$ triplet state, $|T_0\rangle=\frac{1}{\sqrt{2}}(|\uparrow\downarrow\rangle+|\downarrow\uparrow\rangle)$
(the arrows denote the direction of the electron spins).
In singlet-triplet qubits, the exchange interaction is used to drive the $z$-axis rotations
around the Bloch-sphere.
The singlet and triplet states are close to degenerate with zero detuning, and the charge state $|S(1,1)\rangle$ (one electron in
each dot) is the ground state.
Increasing the detuning $\epsilon$ generates an energy splitting between the $S$ and $T_0$ states due to the
exchange interaction. When the detuning is increased enough, the charge state $|S(0,2)\rangle$ becomes
the ground state as the detuning overcomes the Coulomb repulsion caused by occupying one dot with two electrons.
This allows for a projective measurement of the state of the qubit, as the triplet state stays in the $(1,1)$ configuration
due to the repulsive exchange force \cite{petta2}.

In our one-qubit computations, the distance of the parabolic wells of the singlet-triplet qubit is $a=|\mathbf{R}_1-\mathbf{R}_2|=80$ nm. The confinement strength is $\hbar\omega_0=4$ meV.
Other dot-distances and confinement strengths were also studied and the results were qualitatively similar to the ones shown here. We use the 24 first one-particle eigenstates of the system in the many-body ED-computations. This basis size was found to be sufficient for the convergence of
the results (the relative difference of the many-body ground state energies with 18 and 24 single particle states is less than $0.1\%$ up to very high detuning region).

\begin{figure}[!t]
\vspace{0.3cm}
\includegraphics[width=0.5\columnwidth]{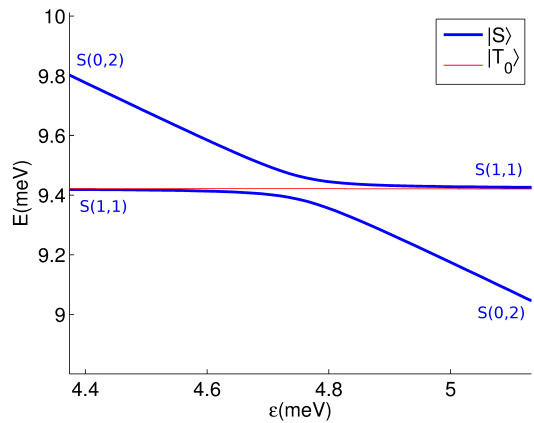}
\centering
\caption{The lowest energy levels of a two-electron DQD-system as function of the detuning. The Singlet states are shown with thick
blue line, and the triplet state ($|T_0(1,1)\rangle$) with the thin red line.
}
\label{fig:ene_2p}
\end{figure}

We first study the energies of the lowest many-body eigenstates as a function of the detuning $\epsilon$. The energy levels are plotted in Fig. \ref{fig:ene_2p}.
The singlet states $|S(1,1)\rangle$ and $|T_0(1,1)\rangle$ are nearly degenerate at low detuning, and $|S(1,1)\rangle$ is the ground state.
Around $\epsilon=4.7$ meV, $|S(1,1)\rangle$ and $|S(0,2)\rangle$ anti-cross, and $|S(0,2)\rangle$ becomes the ground state. In the actual
anti-crossing area, the ground state is a superposition of $|S(1,1)\rangle$ and $|S(0,2)\rangle$. The size of the
anti-crossing gap is $\Delta=68$ $\mu$eV. The transition
from $|T_0(1,1)\rangle$ to $|T_0(0,2)\rangle$ happens at much higher detuning and it is not shown in the figure.

\begin{figure}[!t]
\vspace{0.3cm}
\centering
\includegraphics[width=0.5\columnwidth]{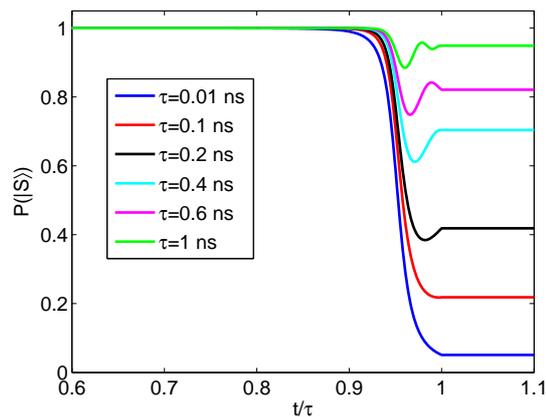}
\caption{The probability of the lowest singlet state ($|S(1,1)\rangle$ with low detuning and $|S(0,2)\rangle$ with high detuning) as a function of normalized time $t/\tau$ with different rise times $\tau$.
The initial state is $|S(1,1)\rangle$. The detuning is increased linearly from 0 to $5.0$ meV in a time of $\tau$. After the detuning has reached its maximum value, the system is
let to evolve for a time of $0.1\tau$. 
}
\label{fig:leak_2p}
\end{figure}

We then study the non-adiabatic charge transitions from $|S(1,1)\rangle$ to $|S(0,2)\rangle$ by sweeping the detuning past the
anti-crossing area to the $(0,2)$-regime (i.e. the the regime where the projective measurement of the singlet probability would be done \cite{petta2}) with varying speeds. The system is initiated in the $|S(1,1)\rangle$ state, and the detuning was then increased linearly
to its maximum value $5.0$ meV during a time of $\tau$. After the detuning has reached its maximum value, the system is let to evolve for a time of
$0.1\tau$. The time step length is $\tau/1000$ (it was found to be small enough to produce accurate dynamics of the situation).
The occupations of the lowest singlet state with different rise times $\tau$ are shown in Fig. \ref{fig:leak_2p}.

When the detuning is increased adiabatically (with respect to the charge state transition) through the anti-crossing area, the occupation of $|S(1,1)\rangle$
in the beginning equals the occupation of $|S(0,2)\rangle$ in the end. If the increase is too fast, the final state is a superposition of $|S(1,1)\rangle$
and $|S(0,2)\rangle$. The faster the increase the bigger the contribution of $|S(1,1)\rangle$. 
The probability of the ground state oscillates at the end of the detuning sweep, as
seen in Fig. \ref{fig:leak_2p}, as changing the detuning couples the two charge states. The oscillations end abruptly when the detuning has reached
its maximum value, at $t/\tau=1$. During the evolution
after $t/\tau=1$, the wave function is a superposition of the eigenstates of the Hamiltonian, and the time-evolution only produces phases, hence
the kink in the probability curves at $t/\tau=1$ (the kink is most prominent in the $\tau=0.2$ ns curve).

Too fast increase of the detuning leads to a Landau-Zener transition to the higher state, $|S(1,1)\rangle$. Indeed, the final probabilities of the
ground state $|S(0,2)\rangle$ agree very well with the Landau-Zener formula,
\begin{equation}
P(|S(0,2)\rangle)=1-\exp\left(-\frac{\pi\Delta^2}{2\hbar |v|}\right),
\end{equation}
where $\Delta$ is the width of the anti-crossing gap, and $v=\frac{d}{dt}(E_{S(0,2)}-E_{S(1,1)})$ is the Landau-Zener velocity. For example with $\tau=0.2$ ns, the simulation (Fig. \ref{fig:leak_2p}) gives $p(|S(0,2)\rangle)=0.42$
at the end of the detuning sweep while the Landau-Zener formula gives $p(|S(0,2)\rangle)=0.43$.

The results are similar if instead of $|S(1,1)\rangle$ the initial wave function is some arbitrary superposition of $|S(1,1)\rangle$ and $|T_0(1,1)\rangle$, i.e. the singlet component
behaves according to the Landau-Zener theory. The same applies also if the initial state is $|S(0,2)\rangle$ and the detuning is decreased to zero linearly.

\begin{figure}[!t]
\vspace{0.3cm}
\includegraphics[width=0.5\columnwidth]{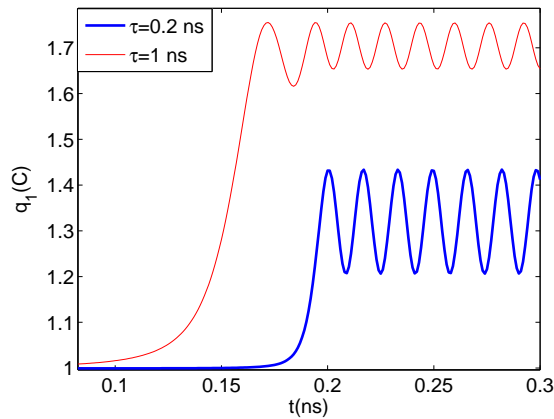}
\centering
\caption{The charge in the left dot of the DQD qubit as a function of time. 
The qubit is initiated in $|S(1,1)\rangle$. The detuning is then increased linearly to $\epsilon=5$ meV in
a time of $\tau$. After the detuning sweep, the system is let to evolve for $0.1$ ns. For clarity, the curves are shown so that in both cases the detuning reaches its maximal value at $t=0.2$ ns
(i.e. the $\tau=0.2$ ns sweep starts at $t=0$ and the $\tau=1$ ns sweep at $t=-0.8$ ns).
}
\label{fig:q_2p}
\end{figure}

In the non-adiabatic case, the final wave function is a superposition of two charge states. The charge density starts to oscillate between the dots
after the detunings have reached their maximal values. Fig. \ref{fig:q_2p}. shows the charge in the left dot (the one with lower potential) during and after the detuning sweep to $\epsilon=5$ meV. 
The charge oscillations are approximately sinusoidal and have the same frequency regardless of $\tau$. As expected, the higher the rise time $\tau$, the more
charge ends up in the left dot (with an adiabatic passage, the charge is constant, $q_1=2$ C after $t=0.2$ ns). The oscillating component of the charge density is rather small compared with the overall amplitude that is determined by
the occupations of the charge states $|S(0,2)\rangle$ and $|S(1,1)\rangle$.

\section{Two qubits}

The logical basis of the two-qubit system (qubits $A$ and $B$) consists of the lowest singlet and $T_0$ states for the two qubits, $\{|SS\rangle,|ST_0\rangle,|T_0S\rangle,|T_0T_0\rangle\}$, where $|SS\rangle=|S\rangle_A\otimes|S\rangle_B$ and so on.
The two-qubit system is simulated as four quantum dots in a line (the minima are located at the $x$-axis).  The four dots are separated into two DQDs. The intra-qubit distance of the minima in the DQDs
is $a_A=|\mathbf{R}_1-\mathbf{R}_2|=a_B=|\mathbf{R}_3-\mathbf{R}_4|=80$ nm. The inter-qubit distance is $|\mathbf{R}_2-\mathbf{R}_3|=120$ nm. The confinement strength is $\hbar\omega_0=4$ meV.

We first study the energy levels of the two-qubit system (four electrons in the four dots) as a function of the detunings $\epsilon_A=V(\mathbf{R}_2)-V(\mathbf{R}_1)$ and $\epsilon_B=V(\mathbf{R}_3)-V(\mathbf{R}_4)$. The 18 first single-particle states were used in the ED computations. The single-particle states
were created with $\epsilon_A=\epsilon_B=4$ meV, as this localizes the states more into the dots with lower potential, which leads to better convergence of the many-body results when the detuning is high. The convergence of the energies was checked, and the basis of 18 states was found
sufficient for good accuracy (the relative difference of the many-body ground state energies with 18 and 24 single particle states was less than $0.1\%$ up to very high detuning region).

\begin{figure}[!t]
\vspace{0.3cm}
\includegraphics[width=0.5\columnwidth]{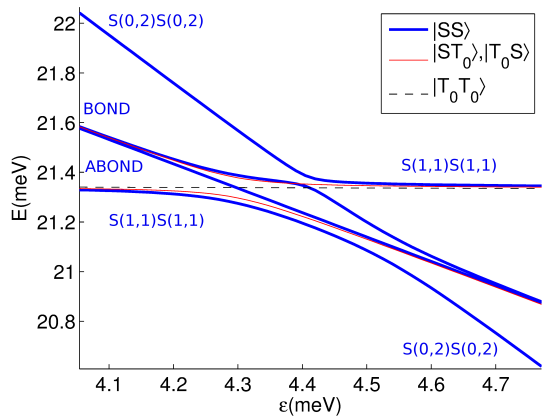}
\centering
\caption{The energy levels of the two-qubit system as a function of the detuning. Both qubits are in the same detuning $\epsilon_A=\epsilon_B=\epsilon$.
The $|SS\rangle$ states are shown with the thick blue lines, the $|ST_0\rangle$ and $|T_0S\rangle$ states with the red lines, and the $|T_0T_0\rangle$ state with the dashed black line.
}
\label{fig:ene_4p_symm}
\end{figure}

The lowest energy levels with symmetric detuning ($\epsilon_A=\epsilon_B=\epsilon)$ are shown in Fig. \ref{fig:ene_4p_symm}. In this case, there is an anti-crossing
area at around $\epsilon=4.4$ meV. It is at a lower detuning compared to Fig. \ref{fig:ene_2p}. due to the fact that the detunings $\epsilon_A$ and $\epsilon_B$ were
defined such that the furthest away dots 1 and 4 are in the low detuning. Hence, the Coulomb repulsion between the dots 2 and 3 facilitates the transition to the $|S(0,2)\rangle$ states.
The more complex anti-crossing region involves four $|SS\rangle$-type states, $|S(1,1)\rangle_A\otimes|S(1,1)\rangle_B$, $|S(0,2)\rangle_A\otimes|S(0,2)\rangle_B$, and two
linear combinations of states $|S(1,1)\rangle_A\otimes|S(0,2)\rangle_B$ and $|S(0,2)\rangle_A\otimes|S(1,1)\rangle_B$. The latter two states are 'the bonding state',
and 'the anti-bonding state',
\begin{equation}\nonumber
\frac{1}{\sqrt{2}}|S(1,1)\rangle_A\otimes|S(0,2)\rangle_B\nonumber
\pm\frac{1}{\sqrt{2}}|S(0,2)\rangle_A\otimes|S(1,1)\rangle_B,
\end{equation}
where $+$ corresponds to the bonding state and $-$ to the anti-bonding state.
Only the bonding state is coupled to $|S(1,1)\rangle_A\otimes|S(1,1)\rangle_B$ and $|S(0,2)\rangle_A\otimes|S(0,2)\rangle_B$.

\begin{figure}[!t]
\vspace{0.3cm}
\includegraphics[width=0.5\columnwidth]{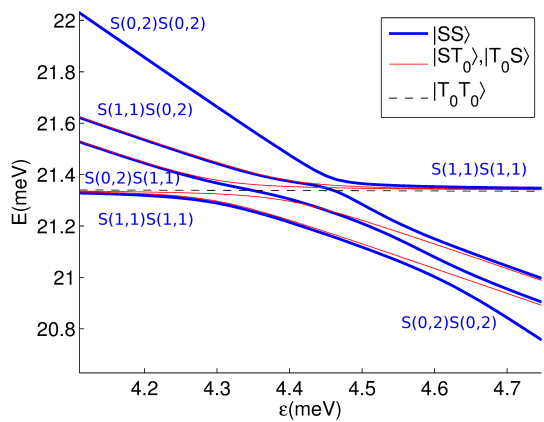}
\centering
\caption{The energy levels of the two-qubit system as a function of the detuning. Here, $\epsilon_B=\epsilon-0.1$ meV, and $\epsilon_A=\epsilon$.
The $|SS\rangle$ states are shown with the thick blue lines, the $|ST_0\rangle$ and $|T_0S\rangle$ states with the red lines, and the $|T_0T_0\rangle$ state with the dashed black line.
}
\label{fig:ene_4p_asymm}
\end{figure}

\begin{figure}[!t]
\vspace{0.3cm}
\includegraphics[width=0.5\columnwidth]{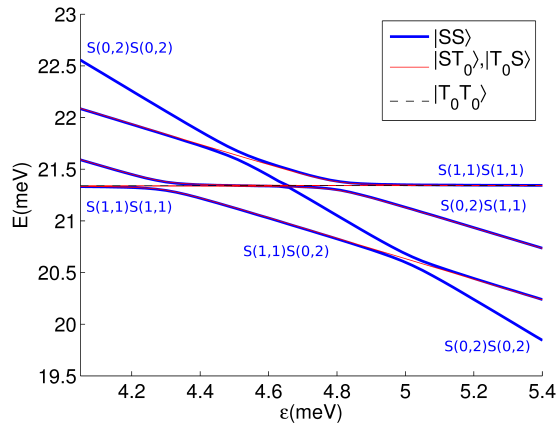}
\centering
\caption{(Color online) The energy levels of the two-qubit system as a function of the detuning. Here, $\epsilon_A=\epsilon-0.5$ meV, and $\epsilon_B=\epsilon$.
The $|SS\rangle$ states are shown with the thick blue lines, the $|ST_0\rangle$ and $|T_0S\rangle$ states with the red lines, and the $|T_0T_0\rangle$ state with the dashed black line.
}
\label{fig:ene_4p_asymm_big}
\end{figure}

Breaking the symmetry of the two qubits splits the bonding and anti-bonding states into $|S(1,1)\rangle_A\otimes|S(0,2)\rangle_B$ and $|S(0,2)\rangle_A\otimes|S(1,1)\rangle_B$.
Fig. \ref{fig:ene_4p_asymm}. shows the energy levels with $\epsilon_A\neq\epsilon_B$. Here, $\epsilon_A=\epsilon$ and $\epsilon_B=\epsilon-0.1$ meV.
As qubit $A$ in now at higher detuning, the corresponding state $|S(0,2)\rangle_A\otimes|S(1,1)\rangle_B$ is at lower energy compared to $|S(1,1)\rangle_A\otimes|S(0,2)\rangle_B$.
 Both states couple to  $|S(1,1)\rangle_A\otimes|S(1,1)\rangle_B$ and $|S(0,2)\rangle_A\otimes|S(0,2)\rangle_B$.
Fig. \ref{fig:ene_4p_asymm_big}. shows
the energy levels with a larger asymmetry $\epsilon_A=\epsilon_B-0.5$ meV$=\epsilon-0.5$ meV. In this case, the detuning difference between the qubits is so large
that there are two anti-crossing regions, one for each qubit. The transition from $|S(1,1)\rangle_A\otimes|S(1,1)\rangle_B$ to $|S(0,2)\rangle_A\otimes|S(1,1)\rangle_B$
happens when $\epsilon=4.8$ meV, and to $|S(1,1)\rangle_A\otimes|S(0,2)\rangle_B$ at $\epsilon=4.3$ meV. The transition to $|S(0,2)\rangle_A\otimes|S(0,2)\rangle_B$ happens at $\epsilon=5$ meV.

\begin{figure}[!t]
\vspace{0.3cm}
\includegraphics[width=0.5\columnwidth]{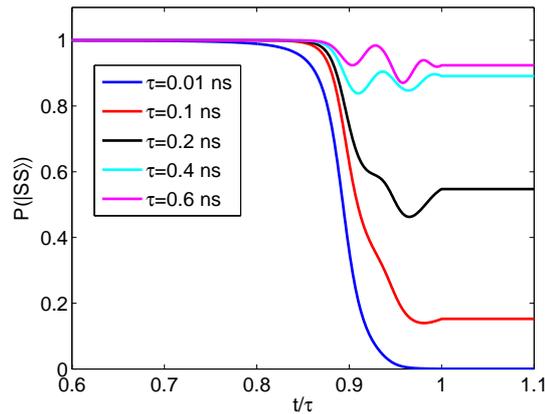}
\centering
\caption{The probability of the lowest $|SS\rangle$ as a function of normalized time $t/\tau$ with different rise times $\tau$.
The initial state is $|S(1,1)\rangle_A\otimes|S(1,1)\rangle_B$. The detuning is increased linearly from 0 to $4.8$ meV in a time of $\tau$. After the detuning has reached its maximum value, the system is
let to evolve for a time of $0.1\tau$.
The probabilities of the ground states shown in Fig. \ref{fig:leak_4p}. look very similar to the one qubit results. However, in the two qubit case, the leakage involves
four charge states $|S(1,1)\rangle_A\otimes|S(1,1)\rangle_B$, $|S(0,2)\rangle_A\otimes|S(0,2)\rangle_B$, $|S(0,2)\rangle_A\otimes|S(1,1)\rangle_B$, and $|S(1,1)\rangle_A\otimes|S(0,2)\rangle_B$
that are all coupled to each other.
}
\label{fig:leak_4p}
\end{figure}

The charge state leakage occurs also in the two qubit system if the detunings are increased too fast. However, as there
are now more states in the anti-crossing region, the phenomenon becomes more complex. Fig. \ref{fig:leak_4p}. shows the probability of the ground state $|SS\rangle$ as a function of time
when the detunings $\epsilon_A$ and $\epsilon_B$ are increased to their maximum values with different speeds. Both qubits are initiated in the $|S(1,1)\rangle$ state, and the detunings are then
increased linearly to their maximum value $4.8$ meV. Here, $\epsilon_A=\epsilon_B$.

With very fast detuning sweeps ($\tau=0.01$ ns and $\tau=0.2$ ns in Fig. \ref{fig:leak_4p}),
the leakage tends to happen mainly to $|S(1,1)\rangle$. A little bit slower increase (still non-adiabatic though) leads to larger occupation
of the bonding state (i.e the states $|S(0,2)\rangle_A\otimes|S(1,1)\rangle_B$ and $|S(1,1)\rangle_A\otimes|S(0,2)\rangle_B$). 
With $\tau=0.01$ ns, the occupations at the end of the simulation shown in Fig. \ref{fig:leak_4p}. are: $P(|S(0,2)\rangle_A\otimes|S(0,2)\rangle_B)\approx0.02$, $P(BOND)\approx0.05$,
and $P(|S(1,1)\rangle_A\otimes|S(1,1)\rangle_B\approx0.94$. With $\tau=0.2$ ns, $P(|S(0,2)\rangle_A\otimes|S(0,2)\rangle_B)\approx0.54$, $P(BOND)\approx0.18$, and 
$P(|S(1,1)\rangle_A\otimes|S(1,1)\rangle_B\approx0.28$. The leakage can happen also between different $|ST_0\rangle$ and $|T_0S\rangle$ charge states. 
In this case, it is effectively similar to the
one qubit case shown in Fig. \ref{fig:leak_2p}, as the $T_0$ part is not affected by the detuning. 

With small asymmetry in the detunings,
the leakage is qualitatively very similar to the symmetric case apart from the fact that the bonding state is now spilt into $|S(0,2)\rangle_A\otimes|S(1,1)\rangle_B$ and $|S(1,1)\rangle_A\otimes|S(0,2)\rangle_B$.
In the highly asymmetrical case in Fig. \ref{fig:ene_4p_asymm_big},
the states $|S(1,1)\rangle_A\otimes|S(1,1)\rangle_B$ and
$|S(0,2)\rangle_A\otimes|S(0,2)\rangle_B$ are not coupled due to the fact that the charge transitions happen at different detunings in the two qubits.
The transitions, and the leakages, happen for one qubit at a time. $|S(0,2)\rangle_A\otimes|S(1,1)\rangle_B$ and $|S(1,1)\rangle_A\otimes|S(0,2)\rangle_B$ leak to $|S(1,1)\rangle_A\otimes|S(1,1)\rangle_B$
at the respective anti-crossings and $|S(0,2)\rangle_A\otimes|S(0,2)\rangle_B$ may leak to $|S(1,1)\rangle_A\otimes|S(0,2)\rangle_B$
at $\epsilon=5$ meV. All these transitions are essentially of the one-qubit type, involving only two charge states, as in Figs. \ref{fig:ene_2p}. and \ref{fig:leak_2p}.

\begin{figure}[!t]
\vspace{0.3cm}
\includegraphics[width=0.5\columnwidth]{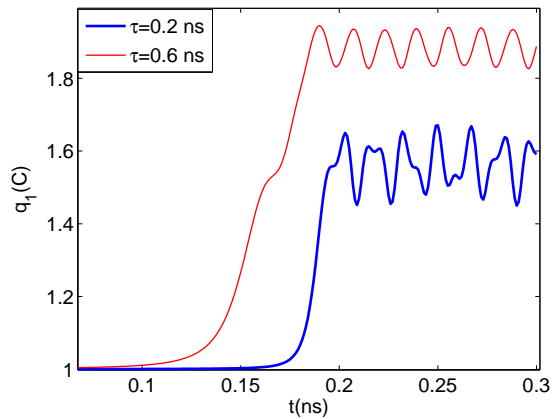}
\centering2
\caption{The charge in the left dot of the qubit $A$ as a function of time. The behavior of the charge in the right dot of the qubit $B$ is identical. The qubits are initiated in
$|S(1,1)\rangle$. The detunings are then increased linearly to $\epsilon_A=\epsilon_B=4.8$ meV in
a time of $\tau$. In the figure, the detuning sweeps end at $t=0.2$ ns. The system is then let to evolve for $0.1$ ns.
For clarity, the curves are shown so that in both cases the detuning reaches its maximal value at $t=0.2$ ns
(i.e. the $\tau=0.2$ ns sweep starts at $t=0$ and the $\tau=0.6$ ns sweep at $t=-0.4$ ns)
}
\label{fig:q_4p}
\end{figure}

Charge oscillations similar to the ones in Fig. \ref{fig:q_2p}. also occur in the two-qubit case. However, the oscillations can be more complex, as there are more
$|SS\rangle$-type eigenstates. Fig. \ref{fig:q_4p}. shows the charge in the left dot of qubit $A$ during and after the detunings are
increased to their maximal values $\epsilon_A=\epsilon_B=4.8$ meV. The charge is plotted only for the qubit $A$ as the behavior of the 
charge in the right dot of $B$ is identical to the one shown. 
The $\tau=0.2$ case exhibits complex charge oscillations as the wave function is now
a superposition of three eigenstates of the system. In the case of $\tau=0.6$ ns, the system is predominantly in the $|S(0,2)\rangle$-state, with much
smaller contribution in the $|S(1,1)\rangle$-state than in the bonding state, $P(|S(0,2)\rangle_A\otimes|S(0,2)\rangle_B)\approx0.914$, $P(|BOND\rangle)\approx0.077$ and $P(|S(1,1)\rangle_A\otimes|S(1,1)\rangle_B)\approx0.006$.
Hence, the oscillations due to $|S(1,1)\rangle_A\otimes|S(1,1)\rangle_B$ are suppressed, and the charge oscillates approximately sinusoidally.

With smaller qubit-qubit distances
(while keeping the intra-qubit dimensions intact), the energy difference between the $|S(0,2)\rangle_A\otimes|S(0,2)\rangle_B$ and the bonding state is increased
due to the stronger repulsion between the qubits. The anti-crossing region also starts at a lower detuning in this case (around $\epsilon=4.15$ compared to
the $\epsilon=4.3$ meV shown in Fig. \ref{fig:ene_4p_symm}).
For example, with the qubit-qubit distance being $100$ nm, there is a $36$ percent increase in this anti-crossing region energy gap compared to the $120$ nm case shown
in Fig. \ref{fig:ene_4p_symm}. The gap between the bonding state and the $(1,1)$-state does not seem to be affected as much. The larger energy gap between the $(0,2)$ state and the bonding state
was found to reduce the charge state leakage slightly. For example, in a sweep to $\epsilon=4.8$ meV in a time $\tau=0.2$ ns, 
the ground state occupation was found to be $P(|SS\rangle)\approx0.60$ compared to the $P(|SS\rangle)\approx0.54$ shown in Fig. \ref{fig:leak_4p}.

The charge state leakage can also affect the functioning of the entangling two-qubit Coulomb gate. 
The qubits A and B can become entangled due to the fact that under the exchange interaction, the charge densities of the $|S\rangle$ and $|T_0\rangle$ states differ (the singlet is a superposition
of the $(1,1)$ and $(0,2)$ charge states), 
and hence the Coulomb repulsion between the qubits depends on the states of the qubits. This conditioning creates an entangled state when the qubits are evolved under exchange.
This allows the creation of a two-qubit CPHASE gate that along with one-qubit operations enables universal quantum computation \cite{taylor, shulman, weperen}.

For the correct operation of the gate, it is necessary to achieve maximal Bell-state entanglement. The degree of entanglement can be determined by some entanglement measure.
One such measure is the concurrence. In this case, it is given as e. g.
\begin{equation}
C=2|\alpha_{SS}\alpha_{T_0T_0}-\alpha_{ST_0}\alpha_{T_0S}|,
\end{equation}
where $\alpha_{SS}=\langle\psi|SS\rangle$, i.e. the projection of the wave function of the system onto the lowest $|SS\rangle$-state, and similarly for the other $\alpha$'s.
Concurrence assumes values between $0$ and $1$. A non-zero $C$ is a property of an entangled state, and the higher the value of $C$, the higher the degree of entanglement. The maximally entangled Bell states have $C=1$\cite{ent}.

We simulate the operation of the two-qubit Coulomb gate. Both qubits are initiated in the $xy$-plane of the Bloch sphere, in the state $|\uparrow\downarrow\rangle=1/\sqrt{2}(|S\rangle+|T_0\rangle)$. The detunings are
then turned on (linearly during a rise time of $\tau$), and when they have reached their maximal values ($\epsilon_A=\epsilon_B=4.28$ meV), the qubits are let to evolve. As the qubits evolve under exchange,
they start to entangle and disentangle in an oscillatory manner. The frequency of the oscillations is proportional to the energy difference $\Delta_E=E_{SS}+E_{T_0T_0}-E_{ST_0}-E_{T_0S}$. 
The concurrence is computed at each time step to study the entanglement of the qubits and the effect of the length of the rise time $\tau$.

\begin{figure}[!t]
\vspace{0.3cm}
\includegraphics[width=0.5\columnwidth]{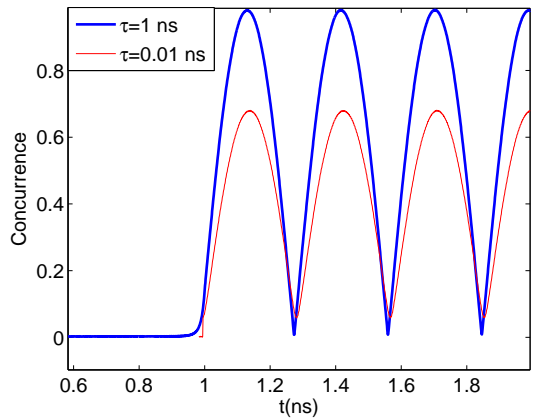}
\centering
\caption{The concurrence as the qubits evolve under exchange. Both qubits, $A$ and $B$, are initiated in the $xy$-plane of
the Bloch sphere. The detunings are then turned on linearly during a time of $\tau$. The thick blue curve shows the adiabatic and the red curve the non-adiabatic case.
(the adiabatic rise started at $t=0$ and the non-adiabatic at $t=0.99$ ns). 
At time $t=1$ ns the detunings have reached their maximal values ($\epsilon_A=\epsilon_B=4.28$ meV) and the qubits are let to evolve for $1$ ns.
}
\label{fig:conc}
\end{figure}

Fig. \ref{fig:conc}. shows the computed concurrence. In the adiabatic case ($\tau=1$ ns), the concurrence reaches its maximal value $1$, i.e. the
two qubit system becomes maximally entangled. When the detunings are increased non-adiabatically ($\tau=0.01$ ns), the concurrence never reaches values above $0.7$ due
to probability leaking out of the qubit basis. The frequency of the concurrence oscillations is not affected by the leakage, as it is determined by the energy difference
of the qubit basis states.

\section{Discussion}

We have studied the non-adiabatic charge state leakage in singlet-triplet qubits using the exact diagonalization method. We have found that when the detuning is increased
too fast, a Landau-Zener transition to a higher lying singlet charge state can occur. In the one qubit case, this transition involves states
$|S(1,1)\rangle$ and $|S(0,2)\rangle$. In the two-qubit case, the situation is more complex. There are four $|S\rangle_A\otimes|S\rangle_B$-type states,
and they all are coupled to each other by detuning. We have also simulated the two-qubit Coulomb gate and studied the effect of the charge state leakage on the gate's
entangling properties. We find out that the leakage can result in the gate not achieving the maximal Bell state entanglement.

Our ED model does not contain
any decoherence effects, as including them would make the computations too heavy. The main source of decoherence in GaAs singlet-triplet qubits is the hyperfine interaction with the semiconductor nuclear spins \cite{khaet,merkulov,folk,Jani1}.  
In $S-T_0$-qubits, the hyperfine interaction couples the singlet to the triplets. It does not affect the coupling between different singlet charge states, that
governs the leakage effect discussed in this paper. In addition, the relevant time scale for the nuclear spin induced decoherence is in the order of tens of nanoseconds (i.e. the dephasing time $T_2^*$
is in this order) \cite{petta2,folk, kopp}, while the leakage effects shown here
become pronounced in the sub-nanosecond scales.

In conclusion, we have found that using too fast detuning pulses can lead to leakage between singlet charge states in $S-T_0$ qubits. This could cause measurement errors
in determining the singlet probability by projecting the state of the qubit onto $|S(0,2)\rangle$, i.e. a singlet could be interpreted as a triplet if the
detuning pulse is too fast. The leakage could also result from quantum gate operation if the gates involve fast detuning pulses, in which case the correct
functioning of the gate could be compromised.

\section*{Acknowledgements}

We acknowledge the support from Academy
of Finland through its Centers of Excellence Program (project no. 251748).

\section*{References}

\bibliography{lagrange}
\bibliographystyle{iopart-num}

\end{document}